\documentclass[twocolumn,preprintnumbers,amsmath,amssymb,floatfix,prd,nofootinbib,showpacs]{revtex4}
\usepackage{amsmath,latexsym, epsfig,
  psfrag,graphicx, rotating, fancyhdr, tabularx, afterpage,booktabs}
\usepackage{dcolumn}
\usepackage{subfigure}
\usepackage{amssymb}
\usepackage{hyperref}
\usepackage{longtable}
\newcommand\nn{\nonumber}                  
\newcommand\ba{\begin{eqnarray}}  
\newcommand\ea{\end{eqnarray}}

\begin{document}
\frenchspacing 
\preprint{ IFJPAN--2013-5 \hskip 1 cm  UAB/FT-731} 
 
\title{ 
Resonance Chiral Lagrangian Currents and 
Experimental Data for  
$\tau^-\to\pi^{-}\pi^{-}\pi^{+}\nu_{\tau}$} 
 
\author{I. M. Nugent}
\affiliation{III. Physikalisches Institut B
RWTH Aachen, D-52056 Aachen, Germany}
\author{T. Przedzi\'nski}
\affiliation{The Faculty of Physics, Astronomy and Applied Computer Science, \\ 
Jagellonian University, Reymonta 4, 30-059 Cracow, Poland}
\author{P. Roig}
\affiliation{Instituto de F\'{\i}sica, Universidad Nacional Aut\'onoma de M\'exico, \\AP 20-364, M\'exico D.F. 01000, M\'exico}
\author{O. Shekhovtsova}
\affiliation{Kharkov Institute of Physics and Technology \\  61108, Akademicheskaya,1, Kharkov, Ukraine}
\affiliation{Institute of Nuclear Physics, PAN,
        Krak\'ow, ul. Radzikowskiego 152, Poland}
\author{Z. W\c{a}s}
\affiliation{Institute of Nuclear Physics, PAN,
        Krak\'ow, ul. Radzikowskiego 152, Poland}
\affiliation{CERN PH-TH, CH-1211 Geneva 23, Switzerland}

\begin{abstract}
In this paper we document the modifications introduced to the previous version of the Resonance Chiral Lagrangian 
current ({\it Phys.Rev.} {\bf D86} (2012) 113008) of the  
$\tau^\pm \to \pi^\pm \pi^\pm \pi^\mp \nu_\tau$ decay which enable the one dimensional distributions measured by 
the BaBar collaboration to be well modeled. The main change required to model the data is the addition of 
the $\sigma$ resonance. 
Systematic errors, theoretical and experimental ones, limitations due to fits of one dimensional distributions only,
 and resulting difficulties and statistical/systematic errors for fitted parameters are addressed.

The current and fitting environment is ready for
  comparisons with the fully exclusive experimental data.  
The present result for $\tau^\pm \to \pi^\pm \pi^\pm \pi^\mp \nu_\tau$ is
 encouraging for work on other $\tau$ decay modes and  Resonance Chiral
 Lagrangian based currents.
\vspace{0.1 cm}
\end{abstract}

\pacs{13.35.Dx, 12.39.Fe, 89.20.Ff, 87.55K-}
\maketitle



\section{Introduction}\label{sec:introduction}

In our paper \cite{Shekhovtsova:2012ra} we described an upgrade of the Monte Carlo generator TAUOLA using 
the results of 
the Resonance Chiral Lagrangian ($R\chi L$) for the $\tau$ lepton decay into the most important two and three meson
 channels. 
The necessary theoretical concepts were collected, 
numerical tests of the implementations were completed and documented. Finally, we presented  strategy for fitting 
experimental data and the systematic uncertainties associated with the experimental measurement. 
However, there was and remain until now,  an obvious limitation due to 
the fact that we are using one-dimensional projections of the invariant masses 
for a multi-dimensional distribution.
The first comparison \cite{Shekhovtsova:2013rb} of the $R\chi L$ results for the $\pi^-\pi^-\pi^+$ mode with the 
BaBar data 
\cite{Nugent:2013ij}, did not demonstrated a satisfactory agreement for the two pion invariant mass distributions.
With the recent availability of the unfolded distributions 
for all invariant masses constructed from observable decay products for this channel \cite{Nugent:2013ij},
we found ourselves in an excellent position to work on model improvement for the $\pi^-\pi^-\pi^+$ mode.
We would like to stress here that the choice of the three pion mode is not accidental. The kinematical configuration is 
complex and the three pion mode 
has the largest branching ratio among the three meson decay modes. Moreover, this decay mode together with the decays 
into two pions, which are much easier to model, are a useful tool 
for spin-parity analysis of the recently discovered Higgs boson \cite{Aad:2012tfa, Chatrchyan:2012ufa} through its 
di-$\tau$ decays \cite{Czyczula:2012ny, Banerjee:2012ez}.

Our paper is organized as follows. In Section 2 we present modifications to the currents previously prepared by 
us~\cite{Shekhovtsova:2012ra}. We give a brief motivation for the choice of selected extensions.
In Section 3, we present numerical results and we discuss available options.
Section 4 is devoted to documentation of our fitting approach, which could be substantially simplified 
thanks to availability of unfolded invariant mass distributions. In Section 
5 we discuss systematic uncertainties for the fit 
resulting from statistical and systematic uncertainties of the experimental data. 
Some technical details of our set-up use are collected 
in this section as well.  
Summary, Section 6, closes the paper.

\section{Extension of currents}\label{sec:ext}

For the final state of three pions $\pi^-$, $\pi^-$, $\pi^+$ with the momenta 
respectively $p_1 $, $p_2 $ and $p_3 $, Lorentz invariance 
determines the decomposition 
of the hadronic current to be \cite{Shekhovtsova:2012ra}
\begin{eqnarray}
J^\mu 
&=&N
\bigl\{T^\mu_\nu \bigl[ (p_2-p_3)^\nu F_1  -  (p_3-p_1)^\nu
 F_2  \bigr]\nonumber\\
& & 
+ q^\mu F_4  -{ i \over 4 \pi^2 F^2}      c_5
\epsilon^\mu_{.\ \nu\rho\sigma} p_1^\nu p_2^\rho p_3^\sigma F_5      \bigr\},
\label{fiveF}
\end{eqnarray}
where:  $T_{\mu\nu} = g_{\mu\nu} - q_\mu q_\nu/q^2$ denotes the transverse
projector, and $q^\mu=(p_1+p_2+p_3)^\mu$ is the momentum of the hadronic system. Here $F$ stands for the pion 
decay constant in the chiral limit. 
In the isospin symmetry limit, the $F_5$ form factor for the three pion mode is zero due to $G$-parity conservation 
\cite{Kuhn:1992nz} 
and thus we will neglect it.
 
Functions $F_i$, the hadronic form factors, depend in general 
on three independent invariant masses that can be constructed from the three meson 
four-vectors. We chose $q^2=(p_1+p_2+p_3)^2$
and two invariant masses  $s_1=(p_2+p_3)^2$, $s_2=(p_1+p_3)^2$ built from 
pairs of momenta. Then
$s_3=(p_1+p_2)^2$ can be calculated\footnote{In our approach the hadronic form factors are calculated in the isospin
limit and
$m_\pi = (m_{\pi^0}+2 m_{\pi^+})/3$.}  from the other three invariants and is
$s_3 = q^2 - s_1 - s_2 + 3m_\pi^2$.  
 The form of the hadronic current is the most general one and constrained 
only by Lorentz invariance. The normalization factor is
 $N = \mathrm{cos} \theta_{\mathrm{Cabibbo}}/F$. 

It is convenient to write down the hadronic form factors as 
\begin{equation}\label{eq:ff_gen}
 F_{i} \ = \ 
 ( F_i^{\chi} \, + \, F_i^{\mbox{\tiny R}} \, + \, F_i^{\mbox{\tiny RR}} )\cdot R^{3\pi}
\ ,\qquad i=1,2,4\ ,
\end{equation}
where $F_i^{\chi}$ is the chiral contribution, $F_i^{\mbox{\tiny R}}$ is the one resonance contribution and 
$F_i^{\mbox{\tiny RR}}$ 
is the double-resonance part. The  $R^{3\pi} $ constant equals 1 for $\pi^-\pi^-\pi^+$ (and -1 for $\pi^0\pi^0\pi^-$). 

The exact form of the function $F_i$ is presented in \cite{Shekhovtsova:2012ra} eqs. (4)-(11). 
The comparison to data \cite{Shekhovtsova:2013rb} hints that the 
lack of the $f_0(600)$ (or $\sigma$) meson contribution in our parameterization may be responsible for 
this discrepancy\footnote{We would like to point out 
that the same problem was shown in the E791 analysis of the $D^+\to\pi^+\pi^+\pi^-$ decays \cite{Aitala:2000xu} (see also the FOCUS article \cite{Link:2003gb}), 
where the $\sigma$ contribution was modeled relying on 
a Breit-Wigner factor. In this case, the inclusion of the $\sigma$ contribution is mandatory to describe the data, 
since it produces basically 
half of the decay width. The good agreement with data shown in \cite{Aitala:2000xu} can be taken as a support of 
modeling the $\sigma$ contribution
by means of a Breit-Wigner, as it is done in this paper, see eq. (\ref{eq:bw_sig}), and can be considered as phenomenologically 
 sufficiently 
sound solution till we have better based parameterization.}.
The $\sigma$ meson is, predominantly, a tetraquark state \cite{Pelaez:2003dy, Caprini:2005zr, Pelaez:2006nj}, and 
it cannot be included in the $R\chi L$ 
formalism\footnote{In addition to the abovementioned $D^+\to\pi^+\pi^+\pi^-$ decays, the $\sigma$ meson has been observed to play an important role in 
$\pi\pi$ scattering \cite{Caprini:2005zr, Yndurain:2007qm, Doring:2009yv}, and it has also been relevant for the understanding of the $J/\Psi\to\omega\pi\pi$ and 
$\Psi(2S)\to J/\Psi\pi\pi$ decays \cite{Beringer:1900zz, pdgsigma}.
However further investigation of its effect in processes with the same hadronic final states, as the one considered in this paper, is required. See our discussion on the fit 
to the experimental data including $\sigma$ and other alternatives, as Coulomb interaction, in section \ref{sec:numeric} and the paper summary.}, which is 
devised for ordinary $q\bar{q}$ resonances%
\footnote{The $q\bar{q}$ assignment for the lightest axial-vector mesons is favored within the $R\chi L$ approach 
\cite{Cirigliano:2003yq}, see however Refs.~\cite{Wagner:2007wy, Geng:2008ag, Wagner:2008gz}.}. 
The inclusion of rescattering effects of 
this kind is an involved task from the computational point of view, even in the case of scalar 
amplitudes \cite{Anisovich:1996tx, Niecknig:2012sj}. In view of this 
we have decided to incorporate the $\sigma$ meson following a more phenomenological approach, more specifically, 
a simple extension  
of the one used by CLEO \cite{Shibata:2002uv}. This effect is included 
into the $F_1(Q^2,s,t)$ and  $F_2(Q^2,s,t)$ form factors in the following way%

\begin{eqnarray}\label{eq:ff_sig}
F_1^{\mbox{\tiny R}} &&\rightarrow  F_1^{\mbox{\tiny R}}+ \frac{\sqrt{2}F_VG_V}{3F^2}
\left[\alpha_\sigma BW_\sigma(s_1)F_\sigma(q^2,s_1) \right. \\
&& \left. + \beta_\sigma BW_\sigma(s_2)F_\sigma(q^2,s_2)\right]\,, \nn \\
F_1^{\mbox{\tiny RR}}&& \rightarrow  F_1^{\mbox{\tiny RR}}+ \frac{4F_AG_V}{3F^2}
\frac{q^2}{q^2-M_{a_1}^2-iM_{a_1}\Gamma_{a_1}(q^2)} \\
&&
\left[\gamma_\sigma BW_\sigma(s_1)F_\sigma(q^2,s_1) + \delta_\sigma BW_\sigma(s_2)F_\sigma(q^2,s_2)\right] ,\nonumber
\end{eqnarray}
where
\begin{eqnarray}\label{eq:bw_sig}
BW_\sigma(x) &=& \frac{M_\sigma^2}{M_\sigma^2 -x -iM_\sigma \Gamma_\sigma(x)} , \\
\Gamma_\sigma(x) &=& \Gamma_\sigma\frac{\sigma_\pi(x)}{\sigma_\pi(M_\sigma^2)} , \label{eq:gamma_sig}\\
F_\sigma(q^2,x) &=& \mathrm{exp}\left[\frac{-\lambda(q^2,x,m_\pi^2)R_\sigma^2}{8q^2}\right] , \label{eq:F_sig}
\end{eqnarray}
and  $\sigma_\pi(q^2) \equiv \sqrt{1 - 4 m_\pi^2/q^2}$ and  $\lambda(x,y,z) = (x - y -z)^2 - 4yz$.
Bose symmetry implies that the form factors $F_1$ and
$F_2$ are related $F_2(q^2,s_2,s_1) =  F_1(q^2,s_1,s_2)$. As a consequence the hadronic current (\ref{fiveF}) 
is symmetrical under exchange of $s_1$ and $s_2$. 
The main differences between the $\sigma$ meson parameterization, eqs. (\ref{eq:ff_sig}) to (\ref{eq:F_sig}), 
and the one used by CLEO can be summarized as:
\begin{itemize}
\item there is one resonance contribution ($F_1^{\mbox{\tiny R}}$) in (\ref{eq:ff_sig}) whereas the CLEO 
parameterization contains  only 
the double resonance contributions ($ F_1^{\mbox{\tiny RR}}$). 
The presence of the one resonance diagrams is a feature of $R\chi L$ (and other Lagrangian approaches), for discussion see Section 7 of  
\cite{Shekhovtsova:2012ra};
\item $\alpha_\sigma \neq \beta_\sigma$ and $\gamma_\sigma \neq \delta_\sigma$, 
no symmetry requirement enforces these equalities in equation (\ref{eq:ff_sig}). In contrary, CLEO Collaboration 
\cite{Shibata:2002uv} uses 
the simplified case where vertices $a_1 \to \sigma \pi$ and $\sigma
\to \pi \pi$ are assumed to be constant and $\alpha_\sigma =
\beta_\sigma$ and $\gamma_\sigma = \delta_\sigma$; 
\item the CLEO Collaboration fixed both the $\sigma$ mass and its width to the values predicted 
by the Tornqvist unitarized quark model \cite{Tornqvist:1995kr} whereas we fit these parameters. 
\end{itemize}

The width for the $\tau^- \to \pi^-\pi^-\pi^+\nu_\tau$ decay can be written down in the terms of the hadronic
 form factors 
in Eqs. (\ref{eq:ff_gen}) and (\ref{eq:ff_sig})
\begin{eqnarray}\label{eq:wid3pi}
\frac{d\Gamma}{dq^2} &=& \frac{G_F^2|V_{ud}|^2}{128(2\pi)^5 M_\tau F^2}\bigg(\frac{M_\tau^2}{q^2}-1\bigg)^2 \\
&&\int ds_1 ds_2 \bigg[ W_{SA} + \frac{1}{3} \bigg(1+2\frac{q^2}{M_\tau^2}\bigg) W_A\bigg] , \nn
\label{q2spec}
\end{eqnarray}
where
\begin{eqnarray}\label{eq:spect_function}
W_{A } &=& - (V_1^\mu F_1 +V_2^\mu F_2 + V_3^\mu F_3 )\\
&& \; \; \; \; \; \; \; (V_{1\mu} F_1 + V_{2\mu} F_2 + V_{3\mu} F_3 )^* \nn \,,\\
W_{SA}&=& q^2 |F_4|^2 \nonumber \,.
\end{eqnarray}
The following phase space integration limits have to be used
\begin{equation}\label{eq:lim_int}
\int ds_1 ds_2 = \int_{4 m_\pi^2}^{\left(\sqrt{q^2}-m_\pi\right)^2} ds_1 \int_{(s_2)_-}^{(s_2)_+} ds_2 \,,
\end{equation}
where
\begin{eqnarray}
(s_2)_{\pm} &=& \frac{1}{4 s}\bigg\{(q^2 - m_\pi^2)^2 \bigg. \\
&-&\bigg. [\lambda^{1/2}(q^2,s_1,m_\pi^2) 
\mp \lambda^{1/2}(m_{\pi}^2,m_{\pi}^2,s_1)]^2\bigg\} \, . \nn
\end{eqnarray}

The Coulomb interaction can be important near the production threshold. We use  
the far-field approximation;
the final-state pions  are treated as stable point-like 
objects and 
the three pion interaction is treated as a superposition 
of the two pion ones. The corresponding invariant masses (we have attractive interaction for 
the pion pairs  of the invariant mases $s_1$ and $s_2$ and the repulsive interaction for the $s_3$)
define the strength of the pair interactions.
The two pion system can be in S or P-wave state. To estimate the Coulomb interaction in S-wave  
we can apply the results of Section 94 of  Ref.~\cite{Landau:101813} which changes 
the differential decay width as follows
\begin{eqnarray}\label{eq:coul}
&&\frac{d \Gamma}{dq^2 ds_1 ds_2} \rightarrow \frac{d \Gamma}{dq^2 ds_1 ds_2}
\frac{2\alpha \pi/v_0(s_1)}{1-\mathrm{exp}\left[-2\alpha \pi/v_0(s_1)\right]}\; \; \; \; \; \; \; \; \;  \\
&&\frac{2\alpha \pi/v_0(s_2)}{1-\mathrm{exp}\left[-2\alpha \pi/v_0(s_2)\right]}\;
\frac{2\alpha \pi/v_0(s_3)}{\mathrm{exp}\left[2\alpha \pi/v_0(s_3)\right]-1 } \nn
\,,
\end{eqnarray} 
where in r.h.s. $d\Gamma$ is the S-wave part of the differential decay width (see eq. \ref{eq:wid3pi}) 
neglecting the Coulomb interactions,
 $v_0(s)$ is the relative velocity of two pions in the two pion system 
with invariant mass $s$, i.e. $v_0(s) = 2\sigma_\pi(s)/(1+\sigma_\pi^2(s))$, where $\sigma_\pi(s)$ is defined below 
eq.~(\ref{eq:bw_sig}). Precision studies of the two-pion vector form factor (see e.g.
\cite{Dumm:2013zh} and references therein), more specifically, a P-wave two pion state, did not
require to include the Coulomb interaction 
between the pion pair to describe the data accurately. We will consequently neglect P-wave Coulomb 
interaction among the final-state pions and stick to formula 
(\ref{eq:coul}) to evaluate the effect of electromagnetic interactions among the pions and set (\ref{eq:wid3pi}) 
on r.h.s. of (\ref{eq:coul}).

The $a_1$ width can be written down as the imaginary part of the two-loop axial-vector--axial-vector correlator with 
suitable flavour indices \cite{Dumm:2009va}
\begin{eqnarray}\label{eq:a1width}
\Gamma_{a_1}(q^2)&=& 2\Gamma_{a_1}^\pi(q^2)\theta\left(q^2 - 9m_\pi^2\right)\\   
&+&2\Gamma_{a_1}^{K^{\pm}}(q^2)\theta\left(q^2 - (m_\pi+2m_K)^2\right) \nn \\
&+& \Gamma_{a_1}^{K^0}(q^2)\theta\left(q^2 - (m_\pi+2m_K)^2\right) , \nonumber
\end{eqnarray}
where 
\begin{eqnarray}\label{eq:a1part}
\Gamma_{a_1}^{\pi,K}(q^2)&& = \\
&&\frac{S}{192(2\pi)^3F_A^2 F^2 M_{a_1}}\bigg(\frac{M_{a_1}^2}{q^2} - 1 \bigg)^2  
\int ds dt W_A^{\pi,K} \nn
\end{eqnarray}
stands for the contribution from the
individual three-pion \cite{Dumm:2009va} and two kaons - one pion \cite{Dumm:2009kj} absorptive cuts. 
As similar integrands are present in eq. (\ref{eq:wid3pi}) for the partial decay width of $\tau$ 
to three pions and in eq. (\ref{eq:a1width}) for the $q^2$-dependent $a_1$ width, we could use 
this property to simplify the code for calculating the invariant mass spectra and  profit in full
from the unfolded invariant mass distribution of  BaBar which recently became public \cite{Nugent:2013ij}.
The Monte Carlo simulations could be avoided and we could use semi-analytical functions in fits. 

We would like to stress  that we neglect, in our numerical analysis, 
the $\sigma$ contribution on 
r.h.s of Eq.~(\ref{eq:a1part}), which should be suppressed in the large-$N_C$ counting.
We will come to discussion of additional ambiguities related to the $\sigma$ contribution in the Summary.

\section{Numerical results}\label{sec:numeric}
\begin{table*}
\begin{tabular}{|l|l|l|l|l|l|l|l|l|}
\hline
      & $M_\rho$   & $M_{\rho'}$& $\Gamma_{\rho'}$& $M_{a_1}$& $M_\sigma$& $\Gamma_\sigma$& $F$& $F_V$ \\
\hline
Min & 0.767 &1.35 & 0.30 & 0.99 & 0.400 & 0.400 & 0.088 & 0.11  \\
\hline
Max &  0.780 &1.50& 0.50 & 1.25 & 0.550 & 0.700 & 0.094 & 0.25  \\
\hline
Fit  & 0.771849 & 1.350000 & 0.448379 & 1.091865 & 0.487512 & 0.700000 & 0.091337 & 0.168652 \\
\hline
\end{tabular}
\begin{tabular}{|l|l|l|l|l|l|l|l|}
\hline
      & $F_{A}$ & $\beta_{\rho'}$ & $\alpha_\sigma$ & $\beta_\sigma$ & $\gamma_\sigma$ & $\delta_\sigma$ & $R_\sigma$ \\
\hline
Min   & 0.1 &  -0.37 & -10.  & -10.  &  -10.   & -10.      &  -10.     \\
\hline
Max   & 0.2 & -0.17 & 10.   & 10.  &  10.     &  10.     &  10.   \\
\hline
Fit   & 0.131425 & -0.318551 & -8.795938 & 9.763701 & 1.264263 & 0.656762 & 1.866913  \\
\hline
\end{tabular}
\caption{Numerical ranges of the $R\chi L$ parameters
used to fit the BaBar data 
for three pion mode \cite{Nugent:2013ij}. 
The approximate uncertainty estimates
are collected in Table \ref{tab:fit-results}.
The $M_{\rho'}$ and $\Gamma_\sigma$ best fit values are observed to be at the boundary of the physically motivated range of variation that we allowed for them.
The $M_{\rho'}$ value may be lower than expected because of missing resonances found in \cite{Shibata:2002uv}. 
On the contrary, $\Gamma_\sigma$ appears to be high because of 
its strong correlation with $R_\sigma$. This correlation, however, does not show up in Table \ref{tab:fitcorr} because its evaluation is not reliable for parameters lying on 
the boundary of the allowed parameter space.}
\label{tab:fit}
\end{table*}

\begin{figure*}
\centering
\includegraphics[scale=.60]{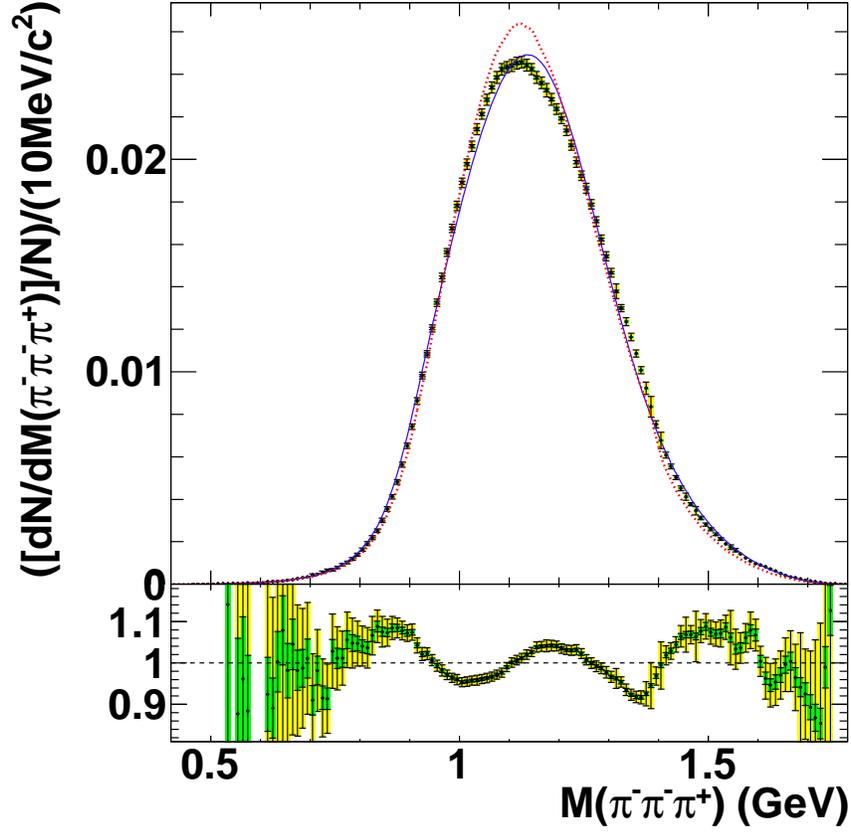}
\caption{The differential decay width of the $\tau^- \to \pi^- \pi^-\pi^+\nu_\tau$ channel is plotted versus the  
invariant mass distribution of the three pion system. The BaBar measurements \cite{Nugent:2013ij} are represented 
by the data points, 
with the results from the $R\chi L$  current as described in the text (blue line) and the
 old tune from CLEO from Refs.~\cite{Golonka:2002iu,Davidson:2010rw} (red-dashed line) overlaid.  
At the bottom of the figure, ratio of new $R\chi L$ prediction to the data is given.
The parameters used in our new model are collected in Table \ref{tab:fit}.  
\label{Fig:pipipi}}
\end{figure*}

\begin{figure*}
\centering
\subfigure{
\includegraphics[scale=.350]{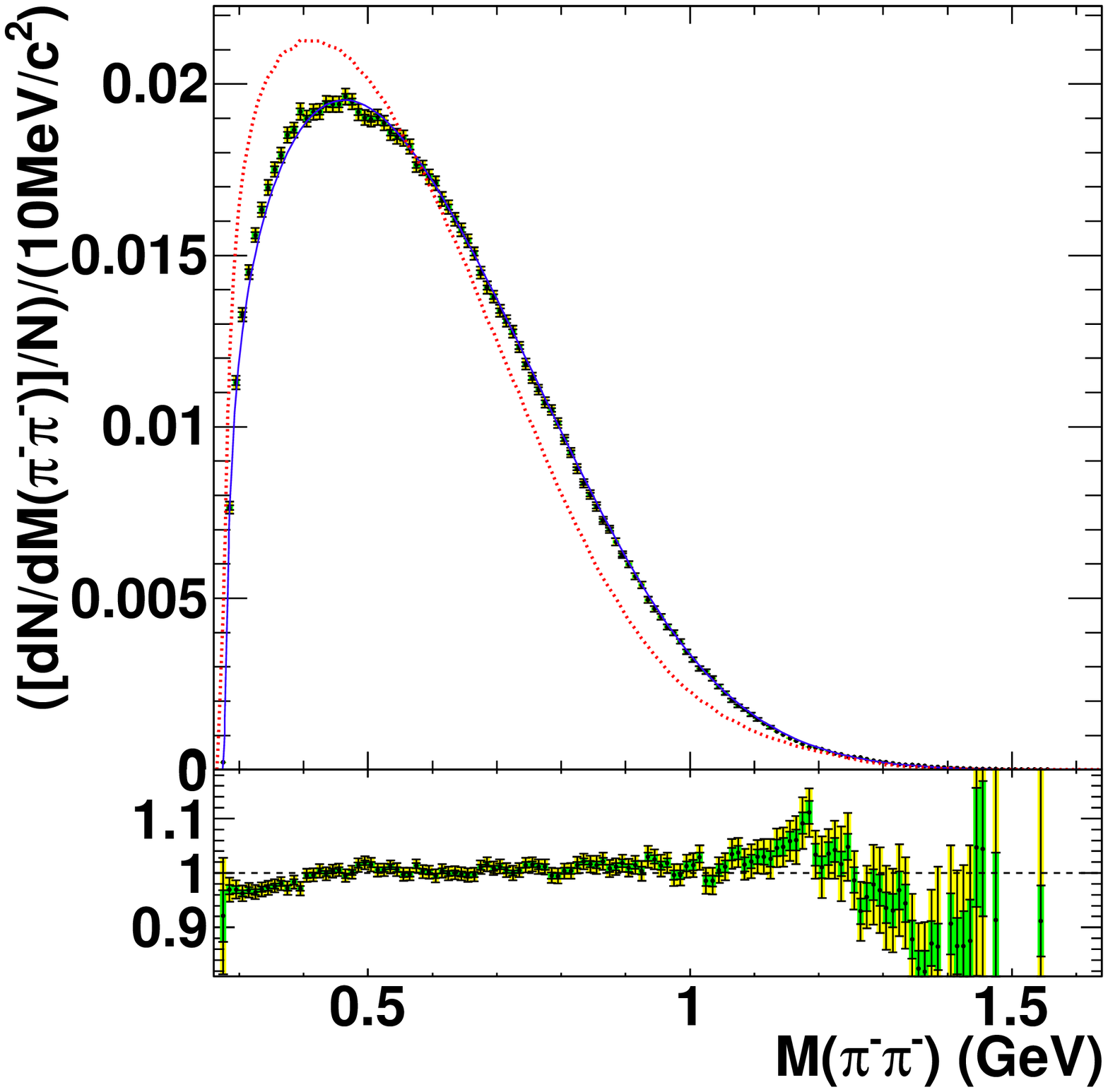}}
\subfigure{
\includegraphics[scale=.350]{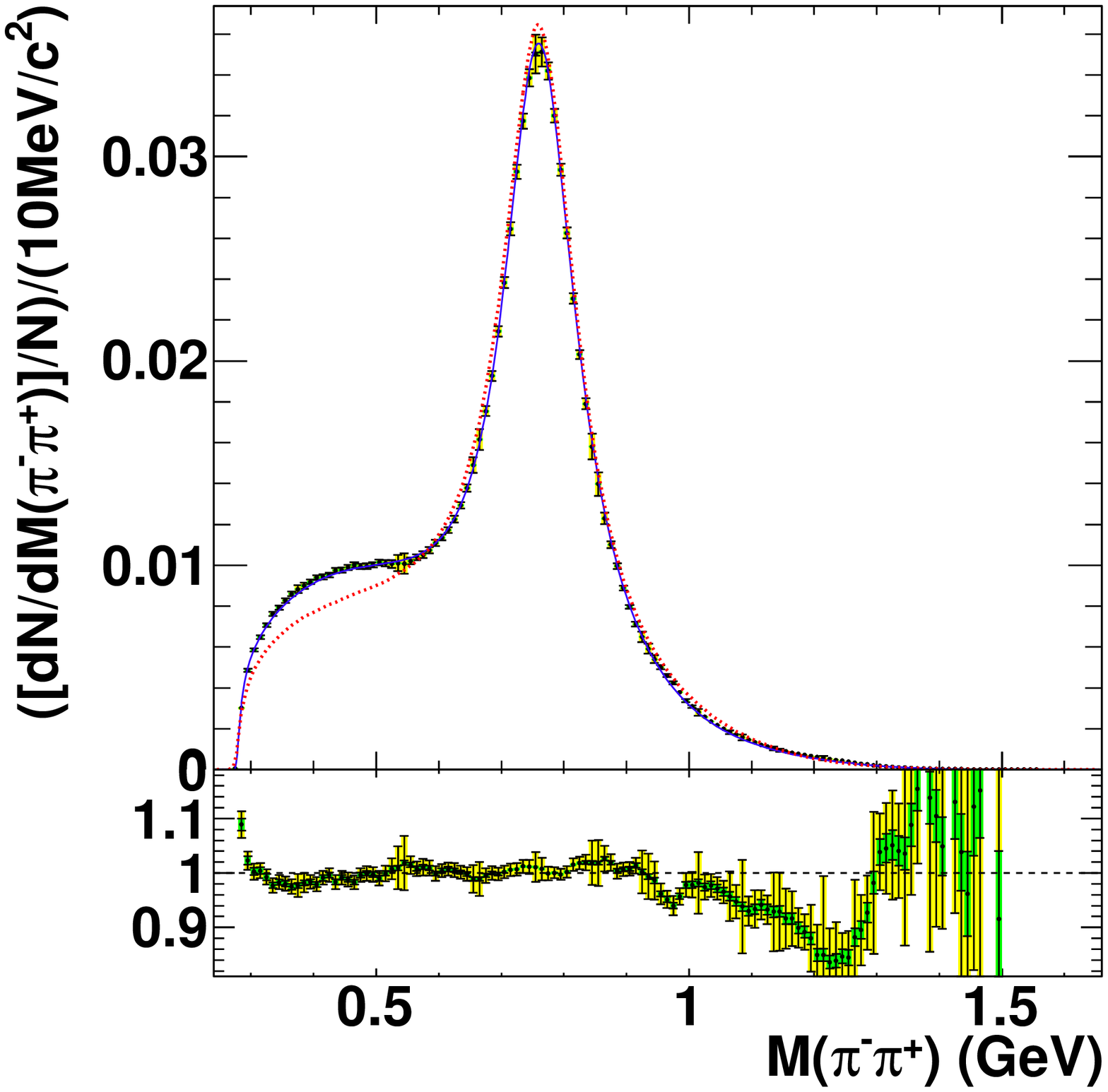}}
\caption{The $\tau^- \to \pi^- \pi^-\pi^+\nu_\tau$ decay 
invariant mass distribution of the two-pion pairs.  The BaBar measurements \cite{Nugent:2013ij} are represented 
by the data points, 
with the results from the $R\chi L$ current as described in the text (blue line) and the
 old tune from CLEO from Refs.~\cite{Golonka:2002iu,Davidson:2010rw} (red-dashed line) overlaid.   
At the bottom of the figures ratio 
of new  $R\chi L$ prediction to the data is given.
The parameters used in our new model are collected in Table  \ref{tab:fit}. 
\label{Fig:pipi}}
\end{figure*}

The parameters of the current%
\footnote{The widths of the $\rho$ and $a_1$ resonances, $\Gamma_\rho\equiv \Gamma_\rho(M_\rho^2)$ and 
$\Gamma_{a_1}\equiv \Gamma_{a_1}(M_{a_1}^2)$, 
are not fit parameters and are calculated by means of eq. (29) in \cite{Shekhovtsova:2012ra} and eq.~(\ref{eq:a1width}).}
 described in the previous Section were used in 
a fit as it is discussed in Section \ref{sec:Fit}. The $\sigma$ contribution is switched on by setting 
{\tt FF3PISCAL} = 2 in {\tt value\_parameter.f}.
The numerical values of the model parameters  are collected 
in Table~\ref{tab:fit}, the goodness of the fit%
\footnote{ In our previous paper, $\chi^2$ was computed using 
the combined statistical 
and systematic uncertainties since only the total covariance matrix was publicly available. For the present results 
we obtain $\chi^2/ndf = 910/401$ when the 
total covariance matrix is used and conditions enabling direct
comparisons are fulfilled.
 This is eight times better than the 
previous result \cite{Shekhovtsova:2013rb}.  The spectra \cite{Nugent:2013ij} used now
and in Ref.~\cite{Shekhovtsova:2013rb} have different binnings. In particular, \cite{Nugent:2013ij} uses a bin width 
of 10 MeV while 
Ref.~\cite{Shekhovtsova:2013rb} used a 20 MeV bin width. Moreover,
the current spectra have smaller uncertainties that are
reduced relative to the results of Ref. \cite{Nugent:2009zz}.}
is quantified by $\chi^2/ndf = 6658/401$. 
We discuss next our best fit numerical results; the plots of invariant masses are given in Fig.~\ref{Fig:pipipi}
and in Fig.~\ref{Fig:pipi}. The partial width resulting from the phase space integration
of the matrix element $\Gamma_{\tau^- \to \pi^- \pi^-  \pi^+ \nu_\tau }=1.9974 \cdot 10^{-13}$ GeV
agrees with the one measured by BaBar    $\Gamma= (2.00\pm 0.03\%)\cdot 10^{-13}$ GeV
 \cite{Aubert:2007mh}  and its PDG value $\Gamma= (2.04\pm 0.01\%)\cdot 10^{-13}$ GeV \cite{Beringer:1900zz}%
\footnote{We quote the branching ratio excluding $\tau^- \to \pi^-\bar{K}^0 \nu_\tau$, with the subsequent decay 
$K^0_S \to \pi^+\pi^-$. The $R\chi L$ current for the three pion mode, 
see Section 2.1 of \cite{Shekhovtsova:2012ra}, does not include the feed down from this mode.
 The value measured by BaBar also excludes the $\bar{K}^0$ contributions. 
The effect of $\tau^- \to \pi^-\omega\nu_\tau$, followed by 
$\omega \to \pi^+\pi^-$ is numerically negligible (branching ratio$\sim3\cdot10^{-4}$) 
and is excluded from our current as well.}.

These results can be confronted with those obtained previously without the $\sigma$ contribution, which can be found 
in Table 2 of \cite{Shekhovtsova:2012ra}.

The effects of the electromagnetic interaction among the final-state pions, see (\ref{eq:coul}), 
is turned on by setting {\tt FCOUL = 1} in {\tt value\_parameter.f}. 
Taking into account the Coulomb interaction (when the $\sigma$ contribution was not included)
we obtained%
\footnote{Note that this result cannot be directly compared with the result from
our previous paper as in this case a different $\chi^2$ function (that takes
into account correlation between histogram bins) is used.}
a $\chi^2/ndf = 33225/401$.
Therefore, the Coulomb interaction without the $\sigma$ contribution cannot describe the data in the low-energy region. 
The effect on the total width is about 2\%, if Coulomb interaction is introduced for currents with
 parameters set as in Table \ref{tab:fit}. When it is introduced at the time of fit the effect is even
smaller.
 It is also negligible for the one-dimensional distributions, the $\chi^2$ changes by 2\% if Coulomb 
interaction is switched on during the fitting procedure,
for both the cases with and without the $\sigma$ contribution.

\subsection{ Case of $\pi^0\pi^0\pi^-$ mode}

A current analogous to eq.(\ref{eq:ff_sig}) can be written for the $\pi^0\pi^0\pi^-$ mode too. 
In this case the neutral pions may be produced by the $\sigma$ meson, exchanged in $s_3$:
\begin{eqnarray}\label{eq:ff_sig_pi0}
&&\!\!\!\!\!F_1^{\mbox{\tiny R}} \rightarrow  F_1^{\mbox{\tiny R}}+ 
\frac{\sqrt{2}F_VG_V}{3F^2}\alpha^0_\sigma BW_\sigma(s_3)F_\sigma(q^2,s_3) \,, \ \\
&&\!\!\!\!\!F_1^{\mbox{\tiny RR}} \rightarrow  F_1^{\mbox{\tiny RR}} \nn \\
&&\!\!\!\!\!\!\!\!+\frac{4F_AG_V}{3F^2}
\frac{q^2}{q^2-M_{a_1}^2-iM_{a_1}\Gamma_{a_1}(q^2)}\gamma^0_\sigma BW_\sigma(s_3)F_\sigma(q^2,s_3)  .\nn
\end{eqnarray} 
The form factors $F_1^{R}$, $F_1^{RR}$ in r.h.s. are the same as for the
$\pi^-\pi^-\pi^+$ mode (up to a common overall sign, see eq.(4) of \cite{Shekhovtsova:2012ra} ) whereas no symmetry
requirement
implies that the $\sigma$ contribution for the $\pi^0\pi^0\pi^-$ decay channel coincides
with the corresponding one in the
 $\pi^-\pi^-\pi^+$ mode. 
We would like to stress that the analysis of CLEO supports the idea that the value of 
the $\sigma$ vertex parameters are different for $\pi^0\pi^0\pi^-$ and $\pi^-\pi^-\pi^+$ modes, 
see Tables 2 and 3 in \cite{Shibata:2002uv}.

To obtain results for $\pi^0\pi^0\pi^-$, we have used the following solution. The
parameters of the $\sigma$: $\alpha_\sigma$, $\gamma_\sigma$, $R_\sigma$, $M_\sigma$ and 
$\Gamma_\sigma$
were fitted again to the $\pi^-\pi^-\pi^+$ data,
varying in the same range as in the default fit, see Table~\ref{tab:fit}.
All three spectra of invariant masses and 
width  were used as experimental 
input to the fit. 
However,  
the additional approximation  $\alpha_\sigma = \beta_\sigma$, $\gamma_\sigma = \delta_\sigma$,
necessary for extension to the $\pi^0\pi^0\pi^-$ mode, was used.
The fit gives\footnote{
Obviously, this constrained fit  for $\pi^-\pi^-\pi^+$ is worse than for the values provided in Table~\ref{tab:fit}
($\chi^2/ndf$ for the new minimum is 21589/401). However, when the new fit is performed with all other parameters 
allowed to vary within the ranges mentioned
in the Table \ref{tab:fit}, 
and only  with the assumption $\alpha_\sigma = \beta_\sigma$, $\gamma_\sigma = \delta_\sigma$,
the corresponding minimum gives $\chi^2/ndf = 8707/401$ which is closer to our default result for 
$\pi^-\pi^-\pi^+$ channel. The results of this fit are
collected in Table~\ref{tab:fit2}. They provide another hint that the $\sigma$ contribution to our decays may be
 considered as a not fully confirmed and/or understood 
phenomenon. }
\begin{eqnarray}\label{eq:fit1}
\alpha_\sigma &=&  1.139486 \, , \;\;  \gamma_\sigma = 0.889769 \, ,  \\
R_\sigma &=& 0.000013\, ,\; \; M_\sigma = 0.550 \, , \; \; \Gamma_\sigma = 0.700 \, . \nonumber
\end{eqnarray}
 Using the results of this fit\footnote{Note that two of the parameters are at the limits we requested for the fit.
 However as we plan to use the results for the $\pi^0\pi^0\pi^-$ mode we leave this
observation without further consideration.},
we may estimate the $\alpha^0_\sigma$ and $\gamma^0_\sigma $ parameters  for the $\pi^0\pi^0\pi^-$ mode 
\begin{eqnarray*}
\alpha^0_\sigma &=&  \alpha_\sigma\cdot Scaling_{factor}^{\gamma}  \\
\gamma^0_\sigma &=&  \gamma_\sigma\cdot Scaling_{factor}^{\gamma} ,
\end{eqnarray*}
 employing the result for the $a_1 \to \sigma\pi$ constant vertex from  Tables 3 and 4 of \cite{Shibata:2002uv}. 
We find $Scaling_{factor}^{\gamma} = 2.1/3.35 = 0.63$. 
Therefore,  for the $\pi^0\pi^0\pi^-$ channel we obtain
$\alpha^0_\sigma = 0.63\cdot 1.139486$, $\gamma^0_\sigma = 0.63\cdot 0.889769$, 
$R_\sigma = 0.000013$, $M_\sigma = 0.55$ and $\Gamma_\sigma = 0.7$.
All other constants remain\footnote{We would like to draw attention to the fact that both fits, this one and the one in Table~\ref{tab:fit2}, 
prefer an $R_\sigma$ value consistent with zero. This coincides with the CLEO result \cite{Shibata:2002uv,Asner:1999kj}.} 
as in Table \ref{tab:fit}; we obtain 
the  $\pi^0\pi^0\pi^-$ partial width\footnote{ 
This was calculated by setting the parameter {\tt BRA1} $= 0$ in routine 
{\tt INITDK}. Results are taken from sample of $2\cdot 10^6$ events.} 
$\Gamma = (2.1211 \pm 0.016\%)\cdot 10^{-13}$, 
$1\%$ higher than the central PDG value and  within the errors cited by PDG.
When all model parameters are adopted in the fit (see Table~\ref{tab:fit2}),
and once the scaling factor is again included, the $\pi^0\pi^0\pi^-$ partial 
width becomes $\Gamma = (2.2706 \pm 0.016\%)\cdot 10^{-13}$ GeV. That is $8.1\%$ higher 
than the PDG value%
\footnote{We note that there is a $\sim 2.4\%$ difference between the $\pi^-\pi^-\pi^+$
and $\pi^- \pi^0 \pi^0$ branching ratios due to phase-space as
a consequence of the $ \pi^0$ mass being smaller than the $\pi^\pm$ mass.}.

The fit results in eq. (\ref{eq:fit1}) reproduce the PDG width of the $\pi^0\pi^0\pi^-$ mode 
better than those in Table \ref{tab:fit2}. 
This is possibly a consequence of accidental partial cancellation of 
phase space effect and  the effect of assumption 
$\alpha_\sigma = \beta_\sigma$, $\gamma_\sigma = \delta_\sigma$ which is not supported theoretically for 
the three charged pion mode. However, it is needed to obtain an estimate of the 
$\alpha_\sigma$ and $\gamma_\sigma$ parameters in the $\pi^0\pi^0\pi^-$ channel for which no data are available.
We  propose the  results of eq. (\ref{eq:fit1}) for the $\sigma$ parameters and for the 
others as given in Table \ref{tab:fit} (i.e. as for the $\pi^-\pi^-\pi^+$ mode) for 
default, until  the $\pi^0\pi^0\pi^-$
experimental distributions or other improvements become available for the fit.
\begin{table*}
\begin{tabular}{|l|l|l|l|l|l|l|l|l|}
\hline
      & $M_\rho$  & $M_{\rho'}$& $\Gamma_{\rho'}$& $M_{a_1}$& $M_\sigma$& $\Gamma_\sigma$& $F$& $F_V$ \\
\hline
Min & 0.767 &1.35 & 0.30 & 0.99 & 0.400 & 0.400 & 0.088 & 0.11  \\
\hline
Max &  0.780 &1.50& 0.50 & 1.25 & 0.550 & 0.700 & 0.094 & 0.25  \\
\hline
Fit  & 0.772774 & 1.350000 & 0.410404 & 1.116400 & 0.495353  &  0.465367  & 0.089675 & 0.167130 \\
\hline
\end{tabular}

\begin{tabular}{|l|l|l|l|l|l|}
\hline
      & $F_{A}$ & $\beta_{\rho'}$ & $\alpha_\sigma$ & $\gamma_\sigma$ & $R_\sigma$ \\
\hline
Min   & 0.1 &  -0.37 & -10.  & -10.  &  -10.        \\
\hline
Max   & 0.2 & -0.17 & 10.   & 10.  &  10.       \\
\hline
Fit   &  0.146848 &  -0.301847 &  1.094981  & 0.582533   &  0.000315  \\
\hline
\end{tabular}
\caption{Numerical values of the $R\chi L$ parameters fitted to BaBar data for three charged pion mode \cite{Nugent:2013ij}
requiring  $\alpha_\sigma = \beta_\sigma$, $\gamma_\sigma = \delta_\sigma$. 
The approximate uncertainty estimates  from {\tt minuit} are
$0.4$ for $R_\sigma$, $0.13$ for $M_{\rho'}$ and  below $10^{-2}$ for rest of 
the parameters.
}
\label{tab:fit2}
\end{table*}

\section{Fitting strategy}\label{sec:Fit}

For the purpose of fitting, a set of semi-analytic distributions over  $s_1$, $s_3$ and $q^2$ 
has been prepared. In the three-dimensional distribution of formula (\ref{q2spec}), two of the parameters 
are integrated over\footnote{ 
For the future application of the fitting procedure
a set of two-dimensional distributions has been prepared as well.}. 
A change of integration variable was introduced to smoothen the integrand and improve integration  convergence.
The 16-point Gaussian integration routine from the old {\tt Fortran CERNLIB} was used. 
Two options were interchanged. In the first one integration adopts the number of divisions for integration domain by 
itself to fulfill the precision requirement. 
In the second  
case the number of divisions has to be provided by the user. Even though the second method
is less sophisticated, it is better suited for fitting algorithms, where minuscule 
changes of integrals due to variation of model parameters are of primary importance.
For all of these methods, a {\tt C} wrappers
have been written and the interface for work with {\tt ROOT} \cite{Antcheva:2009zz} has been prepared.

The $a_1$ width is written down as in eq.~(\ref{eq:a1width}). However, its computation, 
through a double integration for every $q^2$ is demanding on CPU time.
To speed up the calculation, at first, we simply calculated  $a_1$ width as a function of $q^2$ once and we 
kept it in a table, even though it should be re-calculated whenever parameters of the model were changed 
(see Section 4 and Tab. 4  of Ref. \cite{Shekhovtsova:2012ra} for technical details).
This approximation was largely insufficient. Later for  the dominant $3\pi$ contribution to the $a_1$ width 
($\Gamma_{a_1}^{\pi}$ in (\ref{eq:a1width})) 
we followed the method used for a long time since year 1992~\cite{Kuhn:1992nz}: the $a_1$ width 
is calculated by (\ref{eq:a1part}) at several values of $q^2$ only, for other values, it is replaced 
with the polynomial extrapolation  $g(q^2)$ given by eq.~(\ref{GFACT}).  
The coefficients a, b, c, d, e, f, g, h, p, which appear in (\ref{GFACT}), are calculated from the  
set of equations  where the l.h.s. of (\ref{GFACT}) is fixed to the three pion part of the $a_1$ width calculated 
from eq. (\ref{eq:a1width}). 

For the sub-dominant $KK\pi$ contribution to the $a_1$ width a table (see Section 4 and Tab. 4  of Ref. \cite{Shekhovtsova:2012ra} for technical details) is 
 produced, and it is not recalculated with the changing fit parameters 
 at every step of the fit iteration. 
It was checked that the above simplifications did not bring substantial effects for the final results.

We have used the  following form for $g(q^2)$, inspired from Ref.~\cite{Kuhn:1990ad}:
\begin{widetext}
\begin{eqnarray}\label{GFACT}
g(q^2) = \begin{cases} & (q^2-9m_\pi^2)^3(a - b(q^2-9m_\pi^2)+c(q^2-9m_\pi^2)^2), \;\; 9m_\pi^2<q^2<(M_\rho +m_\pi)^2\,,  \\
                       & q^2(d-e/q^2+f/q^4-g/q^6),\hskip 1.5 cm (M_\rho +m_\pi)^2 < q^2 < 3(M_\rho +m_\pi)^2\,,  \\ 
                       & h +2 p\frac{q^2 -3(M_\rho +m_\pi)^2}{(M_\rho +m_\pi)^2} , \hskip 4 cm 3(M_\rho +m_\pi)^2< q^2 < m_{\tau}^2\,, \end{cases} 
\end{eqnarray}
\end{widetext}
where from the best fit we have obtained
\begin{eqnarray}
a&=&1.54712, \; b=3.83256, \; c= 4.52798, \; d=0.30997,\;\nn \\
e&=&1.56106, \; f=3.73605,  \nonumber \\
g&=&2.00856, \; h=0.38688, \; p=-0.00108 \,,\nonumber
\end{eqnarray}
and $M_\rho$ can be taken as the PDG $\rho(770)$ mass and $m_\pi$ as an isospin averaged mass.

As  in Ref.~\cite{Kuhn:1992nz} the expansion in r.h.s. of (\ref{GFACT}) 
starts from $(q^2-9m_\pi^2)^3$ to reproduce the P-wave phase space. 
The difference for  the $q^2$-dependent   $a_1$ width calculated 
as  the sum of (\ref{GFACT}) and the $KK\pi$ table  and the precise results
is less than 7\% starting from $q^2 = 0.293$ GeV$^2$. 
It  decreases below  1\% starting  from $q^2 = 1.1$ GeV$^2$, where it is of major importance. For  $q^2$ 
below $M_\rho^2$ numerical values  for $g(q^2)$ are small, thus of no numerical consequences for the calculated currents.

Once our numerical integration algorithms were understood and the speed of the calculation improved,
 we could   use for the $a_1$ width formula (\ref{eq:a1width})  
and re-tabulate in the manner described in 
Section 4 of ref \cite{Shekhovtsova:2012ra}, whenever parameters of the model were changed in the fit algorithm; parallelization was useful.

In our fits we have used a set of three one-dimensional distributions from BaBar \cite{Nugent:2013ij}. 
We normalize the BaBar data to its branching ratio taken also from the  BaBar measurements \cite{Aubert:2007mh} 
($\Gamma= 2.0\cdot 10^{-13}$ GeV) and perform a $\chi^{2}$ fit using {\tt minuit}
from the {\tt ROOT} library. The covariance matrix provided by BaBar already assumes the statistical uncertainties are Gaussian\footnote{Because the BaBar covariance matrix was constructed
using systematic uncertainties that were estimated with Toy MC, there
are statistical fluctuations that cause instabilities in the matrix
inversion. Therefore, a cut-off on the magnitude of the correlations
was used. The value of the cut-off was determined from where the
$\chi^{2}$, as a function of the cut-off value, becomes unstable.}.  
The $\chi^{2}$ minimized in the fit is determined as the sum of $\chi^{2}$ from 
the three one-dimensional distributions, where correlations between the histograms have been neglected.
 The fitting function has been
 designed in the following way:

\begin{itemize}
\item A set of three histograms is generated using the aforementioned semi-analytic distributions.
\item For each X for which {\tt minuit} requests function value, an appropriate bin content is returned.
\item Whenever {\tt minuit} changes one of the parameters:
\subitem {\tt TAUOLA} is reinitialized with a new set of parameters.
\subitem The $a_1$ width formula is pretabulated\footnote{During first fits, 
at this point the function for interpolation of $a_1$ width (eq. (\ref{GFACT})) was reinitialized.
It was then used instead of $a_1$ table during histogram generation.}.
\subitem A new set of histograms is generated.
\end{itemize}

Without complete retabulation of $a_1$ width, calculating each step took less than a minute on a 2.8GHz processor. It has been further
improved by parallelization of the histogram generation. This has been done by submitting each
bin of each of the three histograms as a separate task distributed evenly among all cores
assigned to the job. This reduced
the time for one step to 6-8 seconds on a 8 core 2.8GHz processor. This approach allows for
flexible assignment of the number of cores used for computation%
\footnote{Note that this method can also be extended to use multiple processors. However,
from our tests the total amount of cores used shouldn't exceed 24. Above this threshold the
communication between the main program and the child processes slows down the whole calculation
giving no gain in time.
This could be solved by introducing more than one main program but we decided against
further optimization of the code as the calculation time was good enough for our purposes.}.

Retabulation
increases computation time only twice. The $\chi^2$ for such a case is about 10\% smaller.
For the final steps of the fits we also include integration over the bin width, instead of previously used
value at the bin center.
This makes $\chi^2$ smaller by another 5-6\% at a cost of 3 to 5 times slower computation.
When both options are included and fits are performed from the point calculated without these
improvements, the $\chi^2$ is reduced yet again by 5\% to 6\%.
Nonetheless such improvements do not change the results in a significant manner, except parameters 
of the $\sigma$. This points to the unconfirmed nature of the $\sigma$ or the insensitivity of the decay 
of the $\sigma$ without the availability of the angular correlations.

Our results collected in section \ref{sec:numeric}, in particular  Figs.~\ref{Fig:pipipi} and \ref{Fig:pipi}, 
have been obtained with both $a_1$ retabulation and integration of bin width taken into account.

\section{Discussion of uncertainties for the fit}

After having completed the presentation of our results, 
rather good agreement 
with the data has been found, substantially better than available in the literature.
Let us turn our attention to the possible systematic uncertainties resulting from the 
use of experimental data available at this moment. 
Our currents used in the fits are not constructed 
posterior to fit properties related to the shapes of the distributions
but are the results of theoretical assumptions. We will cover these points in the following subsections.

\subsection{Numerical Results and Statistical Uncertainties}

The parameters obtained from our fit are collected in Table
\ref{tab:fit-results}. The corresponding correlation matrix can be found in Table \ref{tab:fitcorr}. 
The statistical uncertainties were determined using the {\tt HESSE}
routine from {\tt minuit}~\cite{James:1975dr} 
under the assumption 
that the correlations between distributions and the correlations related to having two entries per event in 
the $\pi^{-}\pi^{+}$ distribution can be  neglected. 
\renewcommand{\arraystretch}{1.5} 
\begin{table*}[h!]
 \begin{center}
\begin{tabular}{|c|c|c|}
\hline
Parameter & number & Value\\
\hline
$\alpha_\sigma$  &  $0$ & $ -8.(795938)  \pm (0.023)  \pm  5$\\
$\beta_\sigma$ & $1$ & $9.(763701)  \pm 0.(013)  \pm 4.0 $\\ 
$\gamma_\sigma$  &  $2$ &$ 1.2(64263) \pm 0.0(09) \pm 0.8$\\ 
$\delta_\sigma$ &  $3$ & $0.6(56762)  \pm 0.0(06) \pm 1.1$\\ 
$R_\sigma$ &   $4$ &  $1.8(66913)  \pm 0.0(053) \pm 1.4$\\ 
$M_\rho$    &   $5$ & $0.7718(49)  \pm 0.0001(8)  \pm 0.0033$\\ 
$M_{\rho^\prime}$  &   $6$ & $1.35(00001)  \pm (9\cdot10^{-6})\pm0.06$
\\ 
$\Gamma_{\rho^\prime}$  &  $7$ & $0.44(8379)  \pm 0.00(6) \pm 0.06$\\
$M_{a_1}$   &   $8$ &$ 1.091(865)  \pm 0.00(029) \pm 0.014$\\
$M_\sigma$   &   $9$ &$ 0.48(7512)  \pm 0.00(033) \pm 0.05$\\
$\Gamma_\sigma$   &  $10$ & $ 0.70(0)  \pm (2.23\cdot10^{-05}) \pm0.21$\\
$F_\pi$ & $11$ & $ 0.0913(37)  \pm (3.3\cdot10^{-5}) \pm 0.0019$\\
$F_V$  & $12$ &$ 0.1686(52)  \pm 0.0001(8) \pm 0.0080$\\
$F_A$  & $13$ &$ 0.131(425)  \pm (6\cdot10^{-5}) \pm 0.011$\\
$\beta_{\rho^\prime}$ & $14$ & $  -0.318(551)  \pm 0.000(9)  \pm 0.034$\\
\hline
\end{tabular}
\caption{The fit parameters presented with the statistical and systematic
  uncertainties. The numbers in round brackets enclose digits that are not significant according to the (other) errors. \label{tab:fit-results}
}
\end{center}
\end{table*}
\begin{table*}[h!]
 \begin{center}
\begin{tabular}{|c|c|c|c|c|c|c|c|c|c|c|c|c|c|c|c|}
\hline
                       & $\alpha_\sigma$ & $\beta_\sigma$ & $\gamma_\sigma$ & $\delta_\sigma$ & $R_\sigma$ & $M_\rho$ & $M_{\rho^\prime}$ & $\Gamma_{\rho^\prime}$ & $M_{a_1}$ & $M_\sigma$ & $\Gamma_\sigma$ & $F_\pi$ & $F_V$ & $F_A$ & $\beta_{\rho^\prime}$ \\ \hline
$\alpha_\sigma$        &    1  & 0.60 & 0.36 &-0.29 &-0.41 &-0.69 & 0.46 & 0.68 &-0.77 &-0.09 & 0.02 & 0.78 & 0.76 & 0.52 &-0.78 \\
$\beta_\sigma$         &  0.60 &   1  & 0.44 &-0.39 &-0.42 &-0.75 & 0.55 & 0.79 &-0.89 &-0.16 & 0.04 & 0.89 & 0.88 & 0.58 &-0.88 \\
$\gamma_\sigma$        &  0.36 & 0.44 &   1  &-0.56 &-0.22 &-0.59 & 0.16 & 0.37 &-0.47 &-0.28 & 0.00 & 0.49 & 0.45 & 0.30 &-0.45 \\
$\delta_\sigma$        & -0.29 &-0.39 &-0.56 &   1  & 0.46 & 0.46 &-0.24 &-0.42 & 0.49 & 0.01 & 0.01 &-0.49 &-0.47 &-0.31 & 0.47 \\
$R_\sigma$             & -0.41 &-0.42 &-0.22 & 0.46 &   1  & 0.42 &-0.33 &-0.56 & 0.62 & 0.34 & 0.02 &-0.53 &-0.56 &-0.42 & 0.48 \\
$M_\rho$               & -0.69 &-0.75 &-0.59 & 0.46 & 0.42 &   1  &-0.27 &-0.64 & 0.79 & 0.29 &-0.02 &-0.83 &-0.74 &-0.48 & 0.75 \\
$M_{\rho^\prime}$      &  0.46 & 0.55 & 0.16 &-0.24 &-0.33 &-0.27 &   1  & 0.67 &-0.61 &-0.13 & 0.03 & 0.61 & 0.66 & 0.37 &-0.65 \\
$\Gamma_{\rho^\prime}$ &  0.68 & 0.79 & 0.37 &-0.42 &-0.56 &-0.64 & 0.67 &   1  &-0.88 &-0.24 & 0.03 & 0.86 & 0.88 & 0.57 &-0.88 \\
$M_{a_1}$              & -0.77 &-0.89 &-0.47 & 0.49 & 0.62 & 0.79 &-0.61 &-0.88 &   1  & 0.28 &-0.03 &-0.96 &-0.97 &-0.62 & 0.95 \\
$M_\sigma$             & -0.09 &-0.16 &-0.28 & 0.01 & 0.34 & 0.29 &-0.13 &-0.24 & 0.28 &   1  &-0.02 &-0.30 &-0.29 &-0.20 & 0.30 \\
$\Gamma_\sigma$        &  0.02 & 0.04 & 0.00 & 0.01 & 0.02 &-0.02 & 0.03 & 0.03 &-0.03 &-0.02 &   1  & 0.03 & 0.03 & 0.03 &-0.04 \\
$F_\pi$                &  0.78 & 0.89 & 0.49 &-0.49 &-0.53 &-0.83 & 0.61 & 0.86 &-0.96 &-0.30 & 0.03 &   1  & 0.95 & 0.55 &-0.97 \\
$F_V$                  &  0.76 & 0.88 & 0.45 &-0.47 &-0.56 &-0.74 & 0.66 & 0.88 &-0.97 &-0.29 & 0.03 & 0.95 &   1  & 0.63 &-0.96 \\
$F_A$                  &  0.52 & 0.58 & 0.30 &-0.31 &-0.42 &-0.48 & 0.37 & 0.57 &-0.62 &-0.20 & 0.03 & 0.55 & 0.63 &   1  &-0.56 \\
$\beta_{\rho^\prime}$           & -0.78 &-0.88 &-0.45 & 0.47 & 0.48 & 0.75 &-0.65 &-0.88 & 0.95 & 0.30 &-0.04 &-0.97 &-0.96 &-0.56 &   1  \\
\hline
\end{tabular}
\end{center}
\caption{The statistical correlation matrix from {\tt minuit} for the fit parameters presented to two decimal places.}
\label{tab:fitcorr}
\end{table*}
\renewcommand{\arraystretch}{1.0} 
We find a strong correlation (correlation coefficients moduli bigger than 0.95) 
between the parameters
$$M_{a_1} , F_\pi, F_V , \beta_{\rho'} .$$ There are also large correlations (with coefficients larger than 0.85)
 between these parameters and $\beta_\sigma$ and 
$\Gamma_{\rho^\prime}$. The correlations between these two last parameters and the former set with $\alpha_\sigma$ and $M_\rho$ 
are only slightly smaller. The 
$\Gamma_\sigma$ is uncorrelated because the Hessian Matrix is not computed correctly for parameters that have a minimum on 
the boundary.

Some of the previously commented correlations are related to the underlying dynamics of the process, as we explain 
in the following.

The dominant contribution to the amplitude originates
 from the exchange $a_1 \to (\rho; \rho')\pi$. 
The explicit form of 
the form factors $F_i^{RR}$ $(i=1,2)$ is given in eqs.~(6-11) 
of Ref.~\cite{Shekhovtsova:2012ra}. 
Since in eq. (\ref{eq:ff_sig}) we use $G_V = F_\pi^2/F_V$, see discussion in Section 7 of Ref. 
\cite{Shekhovtsova:2012ra}, the factor 
multiplying the $a_1$ propagator\footnote{The 
$1/F_\pi$ factor was factored out in the normalization of the currents, see 
eq. (\ref{fiveF}).} is $G_V\cdot F_A/F_\pi^3 = F_V\cdot F_A/F_\pi$ .
As a consequence, strong correlations between $F_V$, $F_A $, $F_\pi$ and also $M_{a_1}$ and 
$\beta_{\rho'}$ could have been expected, as it is the case for all of them but for 
$F_A$ which shows slightly smaller correlations. It could also be guessed that correlations 
affected the other $\rho-\rho^\prime$ parameters ($M_\rho$ and 
$\Gamma_\rho^\prime$), as are indeed observed; but we could not find a reason why strong correlations 
affect also the $\alpha_\sigma$ and $\beta_\sigma$ parameters but not the 
other parameters entering our model of the $\sigma$ contributions ($\gamma_\sigma$, $\delta_\sigma$ and $R_\sigma$).

As we will see later in the paper, for example in the next sub-section on the discussion 
of systematic errors for our results, there are further correlations
between parameters of the model, which are difficult to understand.
This underlines the importance of the systematic uncertainties of our data. 
The actual position of the minimum obtained in our fit may be to some degree
an artifact and result from systematic effects from the measurements which can be accounted for with
systematic and statistical uncertainties. This is important specially for 
the numerical values of strongly correlated parameters. Before the analysis 
of multi-dimensional distributions is available for us in the future, 
let us review the possible consequences of the systematic uncertainties of the data.

\subsection{Systematic uncertainties of the data}

The experimental systematic uncertainties were taken into account
using Toy MC Studies. This approach was adopted instead of studying
individual uncertainties to demonstrate that
systematic uncertainties for the invariant mass measured by Belle and
BaBar can be easily and correctly incorporated into an analysis. The
disadvantage of this method is that one can not isolate the impact of
individual uncertainties and the uncertainties are assumed to be
Gaussian. These limitations are intrinsic to the invariant mass
spectra made available by these collaborations. The Toy 
MC was generated under the Gaussian assumption using Cholesky Decomposition on the
systematic covariance matrix provided by the BaBar experiment to
include the correlations. The fit was re-run for 100 Toy MC to
estimate the impact on the experimental systematic 
uncertainties.  The
resulting systematic uncertainties are presented in Table \ref{tab:fit-results},
while the correlation matrix for the fit parameters can be read from
Table \ref{tab:sys}.  
From Table \ref{tab:fit-results}, it can be seen that the extraction of the fit parameters is limited by the
systematic uncertainties. 
\renewcommand{\arraystretch}{1.5} %
\begin{table*}[h!]
 \begin{center}
\begin{tabular}{|c|c|c|c|c|c|c|c|c|c|c|c|c|c|c|c|}
\hline
& $\alpha_\sigma$ &
$\beta_\sigma$ &
$\gamma_\sigma$ &
$\delta_\sigma$ & $R_\sigma$ 
& $M_\rho$ & $M_{\rho^\prime}$ &
$\Gamma_{\rho^\prime}$ &
$M_{a_1}$ & $M_\sigma$ &
$\Gamma_\sigma$ & $F_\pi$ & $F_V$ & $F_A$ &
$\beta_{\rho^\prime}$ \\ \hline 
$\alpha_\sigma$  &  $1$ &  $-0.84$ &  $-0.27$ &  $-0.15$ &  $0.62$ &  $0.00$ &  $0.47$ &  $0.05$ &  $0.56$ &  $0.17$ &  $-0.61$ &  $-0.05$ &  $0.07$ &  $0.86$ &$0.03$\\
$\beta_\sigma$  &  $-0.84$ &  $1$ &  $0.42$ &  $0.29$ &  $-0.77$ &  $-0.12$ & $-0.37$ & $-0.27$ & $-0.64$ &  $-0.3$ &  $0.83$ &  $0.03$ &  $0.19$ &  $-0.55$ &  $-0.03$\\
$\gamma_\sigma$  &  $-0.27$ &  $0.42$ &  $1$ & $-0.45$ &  $0.01$ &  $-0.47$ &  $-0.24$ & $0.06$ &  $-0.37$ & $0.03$ & $0.23$ & $-0.21$ & $-0.04$ &  $-0.04$ &  $-0.11$\\
$\delta_\sigma$  & $-0.15$ & $0.29$ & $-0.45$ & $1$ & $-0.73$ &  $0.64$ & $-0.15$ & $-0.17$ & $-0.04$ & $-0.68$ & $0.52$ & $-0.26$ &  $0.68$ & $-0.05$ & $0.53$
\\
$R_\sigma$  & $0.62$ &  $-0.77$ &  $0.01$ &  $-0.73$ &  $1$ &  $-0.37$ &  $0.39$ & $0.29$ &  $0.4$ &  $0.5$ & $-0.79$ & $0.21$ & $-0.5$ & $0.39$ &  $-0.32$\\
$M_\rho$  &  $0.00$ &  $-0.12$ &  $-0.47$ & $0.64$ & $-0.37$ &  $1$ & $-0.26$ & $0.05$ &  $0.15$ &  $-0.63$ &  $0.06$ &  $-0.5$ & $0.48$ &  $-0.04$ &  $0.52$\\
$M_{\rho^\prime}$  &  $0.47$ &  $-0.37$ &  $-0.24$ &  $-0.15$ &  $0.39$ & $-0.26$ & $1$ & $-0.38$ & $0.29$ & $0.23$ &  $-0.32$ & $0.35$ & $-0.06$ &  $0.44$ &  $-0.07$\\
$\Gamma_{\rho^\prime}$  &  $0.05$ &  $-0.27$ &  $0.06$ &  $-0.17$ &  $0.29$ & $0.05$ &  $-0.38$ & $1$ & $0.29$ & $0.08$ & $-0.29$ & $-0.24$ &  $-0.28$ & $-0.07$ & $0.06$\\
$M_{a_1}$  &  $0.56$ &  $-0.64$ & $-0.37$ &  $-0.04$ & $0.4$ & $0.15$ & $0.29$ &$0.29$ & $1$ & $0.2$ &  $-0.56$ & $-0.22$ &  $-0.2$ &  $0.55$ & $0.5$\\
$M_\sigma$  & $0.17$ &  $-0.3$ &  $0.03$ &  $-0.68$ &  $0.5$ & $-0.63$ &  $0.23$ & $0.08$ & $0.2$ & $1$ &$-0.32$ & $0.37$ & $-0.67$ & $-0.03$ & $-0.34$\\
$\Gamma_\sigma$  & $-0.61$ & $0.83$ & $0.23$ & $0.52$ &  $-0.79$ &  $0.06$ & $-0.32$ & $-0.29$ & $-0.56$ & $-0.32$ & $1$ & $-0.03$ & $0.44$ & $-0.44$ & $0.19$\\
$F_\pi$ & $-0.05$ & $0.03$ & $-0.21$ & $-0.26$ &  $0.21$ &  $-0.5$ &  $0.35$ &  $-0.24$ &  $-0.22$ & $0.37$ &  $-0.03$ & $1$ &  $-0.18$ & $-0.17$ &  $-0.51$\\
$F_V$   & $0.07$ & $0.19$ & $-0.04$ & $0.68$ & $-0.5$ &  $0.48$ & $-0.06$ & $-0.28$ & $-0.2$ & $-0.67$ &  $0.44$ & $-0.18$ &  $1$ & $0.24$ & $0.5$\\
$F_A$  & $0.86$ & $-0.55$ & $-0.04$ & $-0.05$ & $0.39$ & $-0.04$ & $0.44$ & $-0.07$ &  $0.55$ & $-0.03$ & $-0.44$ & $-0.17$ & $0.24$ & $1$ &  $0.18$\\
$\beta_{\rho^\prime}$ &  $0.03$ & $-0.03$ &  $-0.11$ &  $0.53$ & $-0.32$ &  $0.52$ &  $-0.07$ & $0.06$ & $0.5$ &  $-0.34$ &  $0.19$ &  $-0.51$ & $0.5$ & $0.18$ & $1$\\\hline
\end{tabular}
\end{center}
\caption{The correlation matrix for systematic uncertainties on the fit parameters
  presented to 2 decimal places.}
\label{tab:sys}
\end{table*}
\renewcommand{\arraystretch}{1.0} 
The main systematic uncertainties 
are related to: the limited MC statistics used in the BaBar analysis; the modeling of the detector response for 
the reconstructions of the 
invariant masses, namely the resolution, scale for the momentum and angles of the measured particle as well as 
the bias of the unfolding procedure; and the modeling of the backgrounds. The uncertainty associated with the limited 
MC statistics,
which is comparable to the statistical error on the data, is one of the most significant components of the total systematic 
uncertainty. 
The other main systematic uncertainties impact the shape of the invariant mass spectra. 
The most important one is the detector modeling. As one would expect the uncertainty related to the
modeling of the detector response is most pronounced where the rate of change in the mass spectra is large. This 
uncertainty is of particular importance for determining the width of resonances or parameters correlated to those widths. 
The main backgrounds come from the $\pi^{0}$ fake rate associated with the $\tau^{-}\to \pi^{-}\pi^{-}\pi^{+}\pi^{0}\nu_{\tau}$ 
events
and cross-feed from $\tau^{-}\to K^{-}\pi^{-}\pi^{+}\nu_{\tau}$. 
The $\tau^{-}\to h^{-}h^{-}h^{+}\nu_{\tau}$ cross-feed
from particle mis-identification is primarily
found in the low $Q^{2}$ region from $0.7-0.9$ GeV and the $\pi^{-}\pi^{+}$ region below the $\rho$ resonance.    
The $\tau^{-}\to \pi^{-}\pi^{-}\pi^{+}\pi^{0}\nu_{\tau}$ is most significant around the low $Q^{2}$ region and the 
low $\pi^{-}\pi^{+}$ invariant mass. Since this is the region were the $\sigma$ is added to improve the agreement 
of the fit with data,
the systematic uncertainties are essential to interpret the magnitude of the disagreement and quantifying the 
level at which the $R\chi L$ approach breaks down.
The $q\bar{q}$ background is negligible for most of the invariant mass
spectra except at the high $Q^{2}$ above $1.5$~GeV.
The Bhabha background, although negligible over most of the phase
space, does have a significant contribution to the low $Q^{2}$
region, up to $0.8$ GeV, and the high-$Q^{2}$ region above $1.6$ GeV
as well as below $0.3$ GeV in the $\pi^{-}\pi^{-}$ spectra.
Parameters which are coupled to features that have a 
significant contribution to the spectra, for example the $\rho$ mass,
have systematic uncertainties corresponding to the naive expectation from
the known scale, resolution uncertainties and fit bias uncertainties on
the distribution. 
For other parameters, the conclusions are less clear due to the 
sensitivity in the one-dimensional spectra and strong correlations.
For example, the uncertainties associated with the mass and width of the
$a_{1}(1260)$ and $\rho'(1450)$ are dependent on $\rho(770)$ mass, width and
coupling parameters.
 However, it should be noted that these parameters are likely biased
 and will not correspond to physically meaningful values
 due to missing resonance structures
that have been reported in \cite{Shibata:2002uv,Asner:1999kj}.

\subsection{Test of the convergence of the fitting procedure and Gaussian Approximation}\label{sec:Fittest2}

To verify that our fitting procedure does not depend on the starting 
choice of numerical values for the parameters, and
that it converges properly to a single minimum, we have ran fits from several
different starting points. The procedure of this test is as follows:
\begin{itemize}
\item A random scan of the parameter space has been performed: sample of 210K
      random points has been tested and the $\chi^2$ of the difference between
      these points and data have been stored\footnote{
      Since for this purpose the size of the sample is more important than the
      quality of the result, we were able to make use of the interpolation formula (\ref{GFACT})
      (described in Section \ref{sec:Fit}) instead of dense pretabulation of $a_1$ width.
      We have also neglected integration over bin width. These simplifications introduced
      bias for $\chi^2$ calculation which only weakly depends on the actual choice of the 
      model parameters. However, thanks to this, we have gained a factor of
      7 in speed of calculation, which greatly increased the sample size.}.
\item The sample have been sorted by the $\chi^2$ and a thousand best results
      have been selected.
\item From these results, 20 points have been selected in a way that maximizes
      the distance between these points.
\item From these points we perform a full fit to the data using a method that
      takes into account the correlation between bins in the histograms.
\end{itemize}

We have observed that more than $50\%$ of the results converge to the minimum
described in this paper\footnote{In other cases either {\tt minuit} does
not converge due to number of parameters being at their limit or converges
to other local minima. This verifies our assumption that the fitted function
has several minima. It does, however, point out that the minimum described
in this paper has a very high chance of being a global minimum of this function
as all other minima have distinguishably higher $\chi^2$.}. This is an important 
test that indicates our result are stable and indicates that our fitting 
procedure converges properly regardless of the choice of the starting point.

The Gaussian approximation used for the statistical uncertainty is validated 
using two toy MC studies. For one of the tests we have used the parameter 
values for the current minimum
of the fit to generate 8 Monte Carlo samples of 20 MEvents. We then proceed to
fit our currents to these samples as if they were experimental data samples
with the errors scaled to match the statistics in the data.
The first test started the fit from the very same values used for
generation of the samples. This is to check how our fitting procedure behaves
close to the minimum and how the statistical error of the experimental sample
affects the result. The results are consistent with the starting values.
In the second test, we used the default starting values from our previous
paper~\cite{Shekhovtsova:2012ra}. The result of the fit is in good agreement 
with the values used for generation of Monte Carlo samples.
This indicates that the errors are within the regime were the Gaussian approximation 
for the uncertainties is valid. We assume that any bias from the correlations in negligible.

\subsection{Pole position of the $\rho$, $\rho^\prime$ and $\sigma$ resonances}\label{sec:Polesigma}
We presented our best fit results for the $R\chi L$ parameters with statistical and systematic uncertainties in Table \ref{tab:fit-results}. It may be desirable, however, to 
give the corresponding resonance pole mass and width parameters, which are model independent, to allow for comparisons with other works and encode the spectrum features in 
physically meaningful parameters. We are aware, though, that the complex dynamics of the process under study prevents a competitive determination of the $\rho$, $\rho^\prime$ 
and $\sigma$ resonance parameters, whose pole positions can be determined with higher accuracy in $\pi\pi$ scattering and the vector two pion form factor, the latter either 
in $e^+e^-$ scattering or $\tau$ decays.

Specifically, in order to obtain the corresponding physical mass and width one should compute the position of the pole in the second Riemann sheet of the complex $s$ plane, 
say $s_\mathrm{pole}$. One has (see Refs.~\cite{Bhattacharya:1991gr, Bernicha:1994re, Escribano:2002iv} for the prescriptions on how to deal with the cuts in the complex 
functions entering the corresponding propagators)
\begin{equation} \label{spole}
\sqrt{s_\mathrm{pole}}\, =
\,M_\rho^\mathrm{pole}-\frac{i}{2}\Gamma_\rho^\mathrm{pole}\ .
\end{equation}

If this procedure is carried on for the $\rho$, $\rho^\prime$ and $\sigma$ resonances \footnote{The corresponding calculation for the $a_1$ resonance is beyond the scope of 
this paper, since the relevant three-meson cuts include the $\rho$ and $\rho^\prime$ widths under a double integration making this determination rather cumbersome.} the 
results displayed in Table \ref{Tab:PoleParameters} are obtained.

\renewcommand{\arraystretch}{1.5} 
\begin{table*} 
\begin{tabular}{|l|l|l|l|l|}
\hline
Resonance & $R\chi L$(GeV) Mass & $R\chi L$ Width (GeV) & Pole Mass (GeV) & Pole Width (GeV) \\
\hline
$f_0(500)$ & $0.48\pm0.05$ & $0.70\pm0.21$ & $0.57\pm0.09$ & $0.61\pm0.16$\\
$\rho(770)$ & $0.772\pm0.003$ & $0.149\pm0.007$ & $0.760\pm0.004$ & $0.157^{+0.007}_{-0.009}$\\
$\rho(1450)$ & $1.35\pm0.06$ & $0.44\pm0.06$ & $1.24\pm0.05$ & $0.37\pm0.05$\\
\hline
\end{tabular}
\caption{$R\chi L$ and pole resonance masses and widths (in GeV) for the $f_0(500)$, $\rho(770)$ and $\rho(1450)$ resonances corresponding to our best fit results in 
Table~\ref{tab:fit-results} including statistical and systematic uncertainties. The $R\chi L$ $\rho(770)$ width is given in terms of $F_\pi$ and $M_\rho$ (see Ref.~\cite{Shekhovtsova:2012ra}).}\label{Tab:PoleParameters}
\end{table*}
\renewcommand{\arraystretch}{1.0} 

This kind of analysis is mandatory for the $f_0(500)$ and $\rho(770)$ resonance, for which the determination of its pole parameters has become an area of precision calculations. 
In particular there is very nice agreement in the literature for the $\rho(770)$ pole parameters: $M_\rho^{pole}=761\pm3$ MeV and $\Gamma_\rho^\mathrm{pole}=145\pm4$ MeV using 
different types of data and methods (see \cite{Dumm:2013zh} and references therein for details). Our value in Table \ref{Tab:PoleParameters}, $s_\rho^\mathrm{pole}=
\left(760\pm4-\frac{i}{2}(157^{+7}_{-9})\right)^2$ MeV$^2$ is compatible both for the mass and width, although the last value appears slightly large. We believe that this may be the result of the degeneracies 
and instabilities of the fit caused by the model used for the $\sigma$. In the case of the $\rho(1450)$ resonance, the comparison of our results in 
Table \ref{Tab:PoleParameters} with those in Ref.~\cite{Dumm:2013zh} show agreement for the width but some tension on the mass that is $20$ MeV smaller in Table 
\ref{Tab:PoleParameters}. It is important to note here, that there are observed resonances, $f_{2}(1270)$ and
$f_{0}(1370)$, which were reported CLEO \cite{Shibata:2002uv,Asner:1999kj} and are suggested by the fit to the BaBar data. 
These additional resonances and the correlations between the $\rho$ and $\rho^\prime$ parameters in the fits may also 
have affected the pole parameters of these resonances. Therefore, until  these problems have been solved, 
the pole values for the masses and widths for the $\sigma$, $\rho$ and $\rho^\prime$ may be of limited use.

After a long-standing debate about the existence of the $\sigma$, it is well-established now and the state-of-the-art dispersive analyses of $\pi\pi$ scattering data \cite{Caprini:2005zr, 
Doring:2009yv, Moussallam:2011zg, GarciaMartin:2011jx} agree within a few MeV for the pole position of the $\sigma$ around $((459^{+25}_{-15})-i(279\pm30))$ MeV. Although with 
less precision, the results obtained using $D^+\to\pi^+\pi^+\pi^-$ data, $s_\sigma^\mathrm{pole}\sim (0.47-i0.22)^2$ GeV$^2$ \cite{Oller:2004xm}, are in reasonable agreement with 
the previous values, although with a smaller width. Our figures in Table \ref{Tab:PoleParameters}, $s_\sigma^\mathrm{pole}=\left(0.57\pm0.09-i(0.31\pm0.08)\right)^2$ GeV$^2$, 
are in accord with the more precise determinations obtained analyzing $\pi\pi$ scattering data. Despite this result supports the consistency of our picture where the role of 
the $\sigma$ in the examined decays is manifest (specially in the decay distribution at low values of the $\pi^+\pi^-$ invariant mass) we judge that it is still too early to 
claim this as a confirmed fact and we would like to see if this trend is kept analyzing the complete multi-dimensional data set and/or improving the model of the 
$\sigma$ contribution and possibly adding the missing, for example the $a_1(1640)$ resonance exchange, as it is discussed in the conclusions. 
We are thus confident that in future stages of our 
work, when multi-dimensional distributions will be available for fits we shall be able to improve the determination 
of the poles associated to the resonances whose exchange dominates the $\tau^-\to(\pi\pi\pi)^-\nu_\tau$ decays.

\subsection{Technical details}\label{sec:Technical}

For the above applications to be possible we had to adapt {\tt TAUOLA} library.
Let us document the necessary changes.

The following {\tt FORTRAN} functions with one, two and three-dimensional distributions have been prepared.
Note that all of these functions have been also defined in \\ {\tt demo-fit/wid3pi$\_$demo.h}
for use in C/C++ environment. All of the parameters and return values of these functions
are {\tt DOUBLE PRECISION}.

\begin{itemize}
\item {\tt FUNCTION FFWID3PI(QQ,S1,S3)} \\
      Input: {\tt QQ} = $m_{\pi- \pi- \pi+}^2$, {\tt S1} = $m_{\pi- \pi+}^2$,
      {\tt S3} = $m_{\pi- \pi-}^2$.
      Returns $d\Gamma(\tau^- \to \pi^-\pi^-\pi^+ \nu_\tau)/(d{\mathrm QQ} d{\mathrm S1} d{\mathrm S3})$.
      If QQ S1 S3 are outside of the phase space, this function returns zero.
\item {\tt FUNCTION DGAMS3QQ(QQ)} \\
      {\tt FUNCTION DGAMS1QQ(QQ)} \\
      Calculates $\tau^- \to \pi^- \pi^- \pi^+ \nu_\tau$ width as function of f(QQ,S3) and f(QQ,S1).
      The second parameter of the calculation is hidden in {\tt common block EXTERNAL}.
      The definition in {\tt demo-fit/wid3pi$\_$demo.h} also includes the wrappers
      for these functions that handle the hidden parameter.
\item {\tt FUNCTION DGAMS1(XS1B)} \\
      {\tt FUNCTION DGAMS3(XS3B)} \\
      {\tt FUNCTION DGAMQQ(XQQB)} \\
      These functions calculate $\tau^-\to \pi^- \pi^- \pi^+ \nu_\tau$ width as function of S1, S3 or QQ
      using formula (\ref{eq:ff_gen}), (\ref{eq:ff_sig}), limits of integration are in (\ref{eq:lim_int}).
\end{itemize}

To run the example program, it is required that {\tt ROOT} is installed and its configuration
is available through {\tt root-config}.

\begin{itemize}
\item Compile {\tt TAUOLA-FORTRAN} library by executing {\tt make} in {\tt TAUOLA-FORTRAN/tauola}.
\item Compile additional libraries by executing {\tt make} in {\tt TAUOLA-FORTRAN/demo-standalone}.
\item Compile example by executing {\tt make} in {\tt TAUOLA-FORTRAN/demo-fit}.
\item Execute {\tt ./wid3pi$\_$demo.exe} in {\tt TAUOLA-FORTRAN/demo-fit}.
\end{itemize}

The program should return output: \\
\\
\begin{footnotesize}
{\tt Last integration variable QQ:total width of tau= ...} \\
{\tt Last integration variable S1:total width of tau= ...} \\
{\tt Last integration variable S3:total width of tau= ...} \\
{\tt Calculating h12 (DGAMS3) ...} \\
{\tt Calculating h13$\_$23 (DGAMS1) ...} \\
{\tt Calculating h123 (DGAMQQ) ...} \\
\end{footnotesize}
\\
where the total widths depend on the options and parameters used for computation.
Additionally, output file {\tt out.root}
should be present with the histograms {\tt h12, h13$\_$23} and {\tt h123} containing the distributions
S3, S1 (combined with S2) and QQ correspondingly.

This is a basic example used mostly as a technical test of numerical integration.
It shows how the semi-analytic distributions can be accessed from the C++ program,
and how they can be used to produce the histograms for comparison with the data.
It also shows how fit parameters can be modified.
It can be used as a starting point for further analysis or for writing new fitting algorithm.

The main fitting algorithm is located in subdirectory {\tt TAUOLA/tauola/fitting}.
As the algorithm itself is not a focus of this paper, its detailed
description will be omitted. For users interested in more details we refer
to {\tt README} inside this directory for instructions on how to run the main program
and how to use the algorithms provided. We refer to header files for documentation
of the details of their use. We would like to point out that, as reiterated in
{\tt TAUOLA/fitting/README}, this code is not standalone and requires properly
formatted data files which are not part of the distribution.

Finally let us point that the tar-ball of  software  distribution  which includes code for  the present work 
would also include code for unfinished work on other $\tau$ 
decay modes such as $KK\pi\nu_\tau$. That is why, at present, we plan to distribute the tar-ball
upon request only.

\section{Summary}\label{sec:Sum}  

In the present paper we have documented modification of the Resonance Chiral Theory currents 
for $\tau^- \to \pi^-\pi^-\pi^+ \nu_\tau$ decay of reference \cite{Shekhovtsova:2012ra}, 
necessary for agreement with the experimental data. The choice of this channel was motivated by its
relatively large branching ratio, availability of unfolded experimental distribution and already non-trivial 
dynamics of three-pion final state.  
In addition, this channel is important for Higgs spin-parity studies through the associated di-$\tau$ decays. 
Previously missing contribution from the $\sigma$ resonance was added to our currents and final state 
Coulomb interactions 
were taken into account. As a result, we improved agreement with the data by a factor of 
about eight. Remaining differences are well below uncertainties expected for the  $R\chi L$ currents approach. 
This is the first case when agreement for a non-trivial $\tau$ decay channel was obtained 
between the BaBar data \cite{Nugent:2013ij} and the theoretical model. 
These comparisons allow for future precision tau decays Physics studies \cite{Pich:2013lsa}.

The agreement with data has improved, mainly, due to the inclusion of the 
$\sigma$ meson, which is outside the $R\chi L$ approach. 
In Tables \ref{tab:fit} and \ref{tab:fit-results}, it can be seen that the value of $\Gamma_{\sigma}$ is at the boundary. This appears to be the result
of  a numerical instability in the  phenomenological description of the $\sigma$ caused by the correlation between the
$R_{\sigma}$ in the exponential decay, equation 6, and with width of the Breit-Wigner, equation 4. This problem can
only be resolved with a better parametrization of the $\sigma$ which is currently not available to us%
\footnote{ Parametrization following
 more rigorous lines like in  Refs.~\cite{Anisovich:1996tx,Niecknig:2012sj} 
requires some work on theoretical side and
preferably multi-dimensional distributions for fits.
We postpone such improvement for the next step of the work.}.
Without the
$\sigma$ contribution the $R\chi L$ demonstrates only slightly better agreement with data than the CLEO model
\cite{Shekhovtsova:2013rb}, Fig.1. Therefore, we cannot exclude that fits of similar good quality could be obtained 
if further work
was done on the CLEO model\footnote{It may be interesting to re-do the fits using old models.
Such test may provide more insight on the predictive 
power of $R\chi L$ approach.}. However, the $R\chi L$ approach ensures that
the results for all hadronic currents reproduce the chiral
limit of QCD at least up to next-to-next-to-leading order\footnote{It is
a consequence of the fact that the inclusion of resonances is
done using a Lagrangian, which is built requiring the known
chiral symmetry breaking of QCD, the discrete symmetries of the strong
interaction and unitary symmetry for the resonance fields,
without any ad-hoc dynamical assumption \cite{Ecker:1988te}. Similar
discussion for the 
two pion and $K\pi$ decay modes can be found in Refs.~\cite{Dumm:2013zh, Boito:2010me}.}. In contrast, 
the phenomenological 
approximation, as it was done at CLEO, do not do  that \cite{GomezDumm:2003ku, Coan:2004ep}. 
We will come back to this point in our future analysis of the $KK\pi$ modes.

Even though this work is based on  one dimensional unfolded experimental distributions, 
we think that it is an important step forward. 
We have enriched our fitting arrangements, to profit from the availability of  invariant 
mass distributions. Use of unfolded distributions and our technical arrangements substantially 
improved the speed of our fitting.
We have implemented  solutions based on parallel calculations.

The $R\chi L$ currents are ready for comparison with data for other $\tau$ decay channels and for work 
when unfolded multi-dimensional distributions are used. 
This may constrain the set of parameters obtained from fit to
the $\tau^\pm \to \pi^\pm \pi^\pm \pi^\mp \nu_\tau$ channel, which features strong correlations.
There exist other solutions (local minima) 
for the parameters which give somewhat worse 
 agreement with the experimental data. 
It is worth mentioning that for the results of final fit the {\tt minuit HESSE}
method evaluates some elements of the correlation matrix to surpass 0.9. 

An alternative approach, using linear approximation 
of the dependence on fitted parameters (described in Ref.~\cite{Shekhovtsova:2012ra}),
is pursued independently \cite{Zaremba}. It may become useful in later phases
of the project. It has an advantage that it allows for experimental cuts to be
introduced into the fitted distributions. It is, however, much slower and 
features instabilities if the choice of the fitted function is poor. It provides, however,
independent test of our numerical methods used in the paper.
In the present paper a simpler fitting method has been used as the unfolded
distributions from experimental data are now available \cite{Nugent:2013ij}.

Discrepancies in the high mass region of the $\pi^{+}\pi^{-}$ invariant mass 
indicate the possibility of missing resonances in our $R\chi L$ approach. 
This is consistent with the observation of additional resonances, more
specifically the $f_{2}(1270)$ and
$f_{0}(1370)$,  by CLEO in
\cite{Shibata:2002uv,Asner:1999kj}. Around $975$ MeV, there is another discrepancy between the model and the data
that suggests the presence of the $f_{0}(980)$ resonance\footnote{When looking at the right-hand plot in Fig. \ref{Fig:pipi}, 
the significance of the difference between the model and the data has been computed to be around $4.3$ $\sigma$
in the region from $950$ MeV to $990$ MeV. The $f_{0}(980)$ resonance, as opposed to the $\sigma$, is included in the $R\chi L$ framework \cite{Cirigliano:2003yq} 
and its (small) contribution to the $\tau^-\to\pi^-\pi^-\pi^+\nu_\tau$ decays can with some time be evaluated.}.
Although we could add phenomenologically the contribution of these resonances to the amplitude, 
we prefer not to do it at the moment to keep a compromise between the number of parameters, the stability of the fit and the 
amount of experimental data. We also comment below on the need to include the contribution of the $a_1(1640)$ resonance. 
It should be pointed out that the
necessity of the $\sigma$ in our model may be an artifact caused either by
neglecting the dynamics associated with these additional resonances\footnote{
 The fact that, when $\sigma$ parameters are
unconstrained in the fit, they usually leave the expected physical range 
hints to such a possibility. This is especially highlighted by the fact that even in our results, parameter $\Gamma_\sigma$
is at its upper limit (see Table \ref{tab:fit}). Fit results for $\sigma$ parameters
seem to depend on minuscule changes in its setup, this  
may also support conjecture of unphysical nature of $\sigma$ contribution to the currents or to its inaccurate modeling.}
 or by the crudeness of our $\sigma$ parameterization.  
The largest discrepancies between data and the fitted distribution,
which are responsible for a significant part of the total $\chi^{2}$, 
are observed in the 3$\pi$ invariant mass distributions. 
The slope and shape of the disagreement in the 3$\pi$ invariant mass spectra, in particular around $1.5$ GeV in
Fig. \ref{Fig:pipipi}, indicates the possibility of interference between
$a_1(1260)$ and its excited state $a_1(1640)$. The disagreement in the
low mass regions around $1.0$ GeV and below could be the result of a
bias in the fit caused by leaving out the additional resonance.
 Although we currently restrict ourselves to the lowest axial-vector resonance, $a_1(1260)$, in future analysis 
of multi-dimensional distributions we are going to include the $a_1(1640)$ in the same way
 as it was done for the $\rho'(1450)$ to test our hypothesis\footnote{
Since $M_{a1(1640)}\sim M_{\tau}$ the constant width approximation can be used for this resonance.
 }. We prefer not to do this at the moment because the improvement of the fit on the peak and tail regions in 
the three-meson invariant mass distribution 
will come again together with an increase of the fit instability and a growth of the correlations between 
the (larger number of) fit parameters. Lacking the complete 
multi-dimensional distributions it is advisable to proceed this way at present. 
CLEO has also suggested
a hint of a possible contribution of the pseudo-scalar resonance $\pi'(1300)$, which they  have
 excluded at the $(1.0-1.9)\times10^{-4}$ level with a 90\%
 CL\cite{Shibata:2002uv,Asner:1999kj}. This is below the current
 sensitivity of our fit. Finally, we note that previous dedicated studies within the $R\chi L$ framework 
\cite{Dumm:2009va, GomezDumm:2003ku} 
did not obtain an improved description of the data when including the chiral logarithms \cite{Colangelo:1996hs}. 
This question could also be readdressed 
when more exclusive data will become available.

This paper represents a significant improvement in the agreement between our model 
and the data. The resulting fit parameters are substantially 
improved and are reasonable from the point of view of internal consistency 
of $R\chi L$. The obtained value for the $a_1$ mass can be used for the future theoretical study.

Predictions for the $\pi^0\pi^0\pi^-$ are obtained from the ones for 
$\pi^-\pi^-\pi^+$ directly with the help of isospin symmetry without 
any modifications except the part of $\sigma$ contribution which was adapted 
and constraints from CLEO were used.  
Then, our current results with the partial branching ratio, which 
is $1.1\%$ higher than the 
PDG result, is reasonable for our somewhat simplistic assumption.

 On technical side, we have prepared fitting environment which can be after minor 
adaptation used for multi-dimensional experimental distributions or other $\tau$ decay modes.

\section*{ Acknowledgements}
We acknowledge help and discussion with Jakub Zaremba and Sergey Alekhin on numerical details of fitting arrangements. 
We are also thankful to Alexander Korchin and Pedro Ruiz Femen\'ia   for a discussion about the Coloumb interaction and 
to Rafel Escribano, Bastian Kubis and 
Jorge Portol\'es for their comments about the $\sigma$ meson.
Useful discussions with John Michael Roney, Simon Eidelman, Hisaki Hayashii,
Denis Epifanov  and Swagato Banerjee are acknowledged. 
We thank members of BaBar collaboration for discussions, in particular William Dunwoodie for discussion of the $\sigma$ meson. We 
benefited from discussions with Daniel G\'omez Dumm on the determination of the pole parameters of the resonances exchanged in $\tau^-\to\pi^-\pi^-\pi^+\nu_\tau$ decays.

This project is financed in part from funds of Polish National Science
Centre under decisions  DEC-2011/03/B/ST2/00107 and DEC-2011/03/B/ST2/00220.
We acknowledge funding from the Alexander von Humboldt-Stiftung/Foundation and the Spanish grant FPA2011-25948 as well. 
P.R. acknowledges DGAPA for funding his contract and the support of project PAPIIT IN106913. 
O.S. acknowledges that the final step of this project is financed in part from funds of Foundation of Polish Science 
grant POMOST/2013-7/12. POMOST Programme is cofinanced from European Union, Regional Development Fund. 
This project was supported in part by PL-Grid Infrastructure.


\end{document}